# RADiCAL – Precision-timing, Ultracompact, Radiation-hard Electromagnetic Calorimetry

T. Anderson[4], T. Barbera[1], D. Blend[3], N. Chigurupati[4], B. Cox[4], P. Debbins[3], M. Dubnowski[4], M. Herrmann[3],
C. Hu[2], K. Ford[1], C. Jessop[1], O. Kamer-Koseyan[3], G. Karaman[3], A. Ledovskoy[4], Y. Onel[3], C. Perez-Lara[4],
R. Ruchti[1], D. Ruggiero[1], D. Smith[1], M. Vigneault[1], Y. Wan[1], M. Wayne[1], J. Wetzel[3], L. Zhang[2] and R-Y. Zhu[2]

[1]University of Notre Dame, Notre Dame, IN 46556
[2]California Institute of Technology, Pasadena, CA 91125
[3]University of Iowa, Iowa City, IA 52242
[4]University of Virginia, Charlottesville, VA 22904

**Abstract**
To address the challenges of providing high performance calorimetry in future hadron collider experiments under conditions of high luminosity and high radiation (FCC-hh environments), we are conducting R&D on advanced calorimetry techniques suitable for such operation, based on scintillation and wavelength-shifting technologies and photosensor (SiPM and SiPM-like) technology.  In particular, we are focusing our attention on ultra-compact radiation hard EM calorimeters, based on modular structures (RADiCAL modules) consisting of alternating layers of very dense absorber and scintillating plates, read out via radiation hard wavelength shifting (WLS) solid fiber or capillary elements to photosensors positioned either proximately or remotely, depending upon their radiation tolerance. The RADiCAL modules provide the capability to measure simultaneously and with high precision the position, energy and timing of EM showers.  This paper provides an overview of the instrumentation and photosensor R&D associated with the RADiCAL program.

## 1. Introduction

The R&D objective and goals of RADiCAL are focused on the development of precision EM calorimetry for future hadron colliding beam experiments and are directed toward addressing the Priority Research Directions (PRD) for calorimetry listed in the DOE Basic Research Needs (BRN) workshop for HEP Instrumentation [1].

**Objective:** To construct an array of ultracompact RADiCAL modules to explore the potential of ultracompact EM calorimetry capable of precision timing, energy and position measurements and specialized particle identification in high radiation fields [2]. To reach the objective requires R&D on radiation hard and fast response scintillators, wavelength shifters and fiberoptic elements and photosensors.

**Initial goal:** The initial goal is to establish a performance baseline for RADiCAL modules through instrumentation capable of delivering an EM energy resolution approaching $\sigma_E/E = 10\%/\sqrt{E} \oplus 0.3/E \oplus 0.7\%$ [3], a timing resolution $\sigma_t < 50$ps [4], and position resolution for the shower centroid within a few mm [5].  This initial effort uses radiation hard optical elements, but is instrumented with currently available SiPM photosensors, which are adequate for beam tests to establish and characterize a performance baseline for the RADiCAL technique, but will not be performant if placed in high radiation areas.  Once the time/energy/position baseline is established, R&D will be focused on the further development and refinement of radiation hard optical components and new photosensors to replace those with vulnerabilities.

**Primary goal**: To develop candidate instrumentation capable of operation in the FCC-hh endcap region up to $|\eta| \leq 4$ with an EM energy resolution indicated above, noting that for $|\eta| \leq 2.5$, the environmental conditions are expected to be 100Mrad ionization dose and $3\times10^{16}$ 1MeV $n_{eq}$ fluence [3]. This effort will include further optical materials development and R&D on radiation hard photosensors to replace the conventional SiPM.

**Stretch goal:** To identify future directions and candidate instrumentation with the potential for operation in the FCC-hh in endcap/forward regions, where the operating conditions are foreseen to be a sobering 500 Grad ionization dose and $5\times10^{18}$ 1MeV $n_{eq}$ fluence [3]. To reach and function in this domain will require further creative innovations in optical materials and photosensor development, guided by the work toward the Primary goal.

**Results from Ongoing R&D and Research Plan.** Our research and development plan and testing approach are guided in part by our prior work on the development of radiation hard Shashlik-style technology, a calorimetry technique proven to be effective in the hadron collider environment [3-5] and initially explored by our group in the context of the CMS Phase 2 Upgrade [6]. In the sections below we provide:
- Section 2: Overview of initial modular array tests, that have informed the RADiCAL concept.
- Section 3: Ongoing R&D on component instrumentation for RADiCAL.
- Section 4: Beam testing plans.

**2. Overview of Initial modular array tests.** The initial RADiCAL prototype (Fig. 1) is a Shashlik-style module consisting of a layered structure of 29 LYSO crystal plates interleaved and 28 W plates. Four WLS liquid-filled and sealed quartz capillaries are used to wave-shift the scintillation light and transfer this light to photosensors located either at the downstream end of the module (right hand side in the figure) or via fiberoptics to photosensors positioned remotely. A centrally placed optical fiber running through the length of the module is used to inject pulsed laser light for monitoring.

The LYSO:Ce and W layering have been selected for their excellent performance characteristics. The particular modular structure shown in Fig. 1 has a depth of ~25$X_o$ and a sampling fraction of 18%, properties which could be readily adjusted to greater or lesser values by adding/removing layers and/or adjusting layer thicknesses. For size comparison, eight of these modules could be fit within the volume of a single $PbWO_4$ crystal currently used in the CMS Endcap (EE) calorimeters [7]. Stand alone GEANT4 simulations (see Figs. 2,3) have indicated that this type of technique would provide a stochastic energy resolution ~10%/√E (dominated by sampling fraction), and a constant term of less than 1% (dominated by leakage) and excellent capability for mitigating the effects of event pileup (at 200 interactions per beam crossing) due to the small Molière radius (13.7mm) and compact Modular dimensions of 14 x 14 x 114 $mm^3$ [8,9].

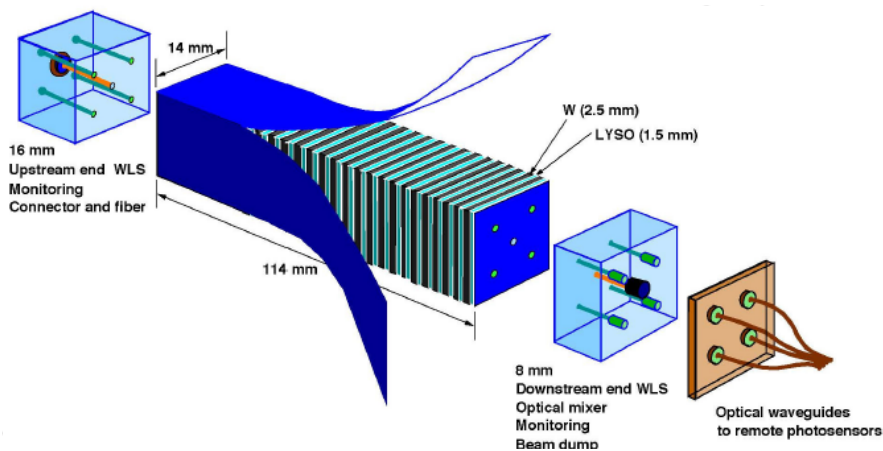

Figure 1. The initial RADiCAL prototype for ultra-compact EM calorimetry based upon interleaved layers of W and LYSO:Ce crystal in a Shashlik-like structure and read out with specialized WLS elements.

The beam tests of a 4x4 array (Fig. 3) were carried out at CERN in the H4 beamline. The active volume of the array was 56mm x 56mm x 114mm, and there were 64 independent readout channels for this structure, with 4 WLS capillaries per module, each capillary connected by a clear fiber waveguide to its own individual HPK SiPM photosensor having 15µm square pixels.

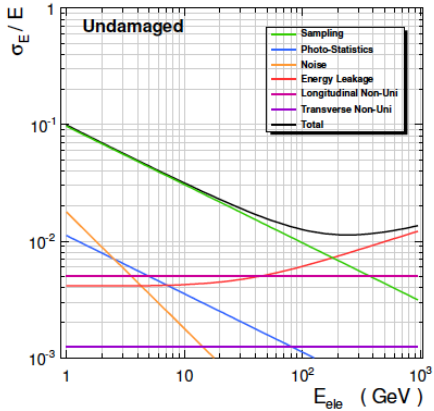
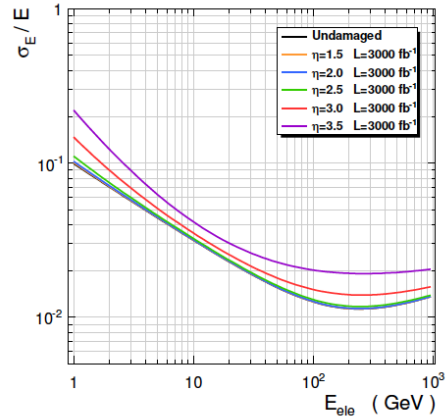

Figure 2. Energy resolution for an un-irradiated Shashlik-style endcap calorimeter built of modules shown in Fig.1, indicating the contributions to the resolution from various sources. GEANT4 simulation.

Figure 3. Evolution of the energy resolution of a Shashlik-style calorimeter, as a function of integrated luminosity and for various regions of eta in the CMS endcap under HL-LHC conditions. Note there is little change to the resolution for $|\eta|$ values below 2.5.

Figure 4 shows the energy resolution achieved with this array, providing a comparison of DSB1 WLS capillary readout and multiclad Y11 WLS fiber readout. The data indicate that the rad-hard capillary methodology provides the same energy resolution as that of the much less radiation hard Y11 fiber technology. Given the high light levels produced by the LYSO and utilizing the fast waveshifter DSB1 in either capillary or fiber form, fast timing capability is possible with these structures based on the integrated optical signal from the full LYSO/W module. Figure 4 also shows the results of a study of the timing resolution as a function of electron beam energy, for readout with both DSB1 WLS organic plastic fiber and DSB1 WLS liquid-filled capillaries [10]. The data indicate that the timing resolution improves with greater detected light levels (faster signal rise time), achieving ~ 60ps for a single fiber or capillary with SiPM readout (here again a HPK device with 15μm square pixels).

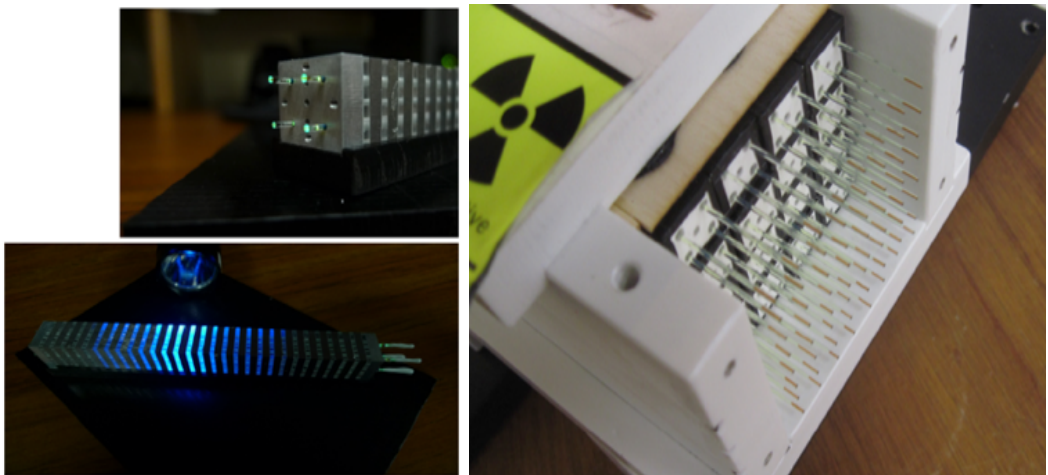

Figure 4. Photographs of the LYSO/W 4x4 module test array used in beam tests equipped with DSB1 WLS capillaries. Left: Structure of an individual Module (left). Right: The 4x4 Modular Array during assembly.

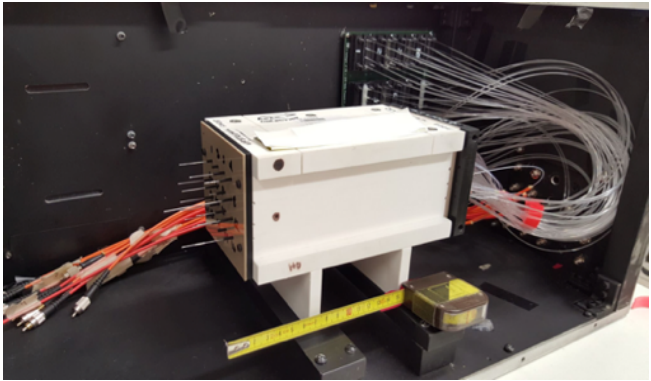
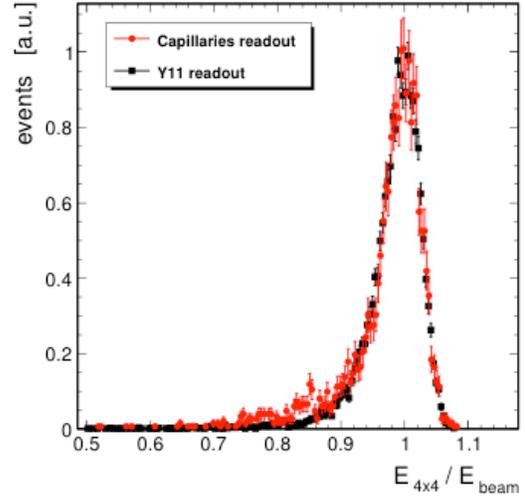

Figure 5. Fiber optic waveguides used to transmit the WLS light to SiPM photosensors positioned outside of the beam region for this specific test. In the picture, the incoming electron beam is incident from the left on the 4x4 modular array enclosed within the white mechanical housing.

Figure 6. Energy resolution for the 4x4 W/LYSO Array using capillary WLS readout out (red) compared to Y11 double clad WLS fiber readout (black), as measured using an electron beam in the CERN H4 beam line. Electron beam energy is 100 GeV.

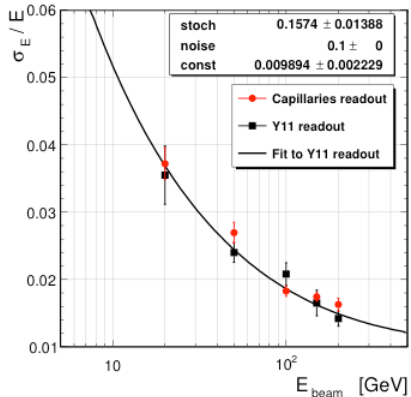
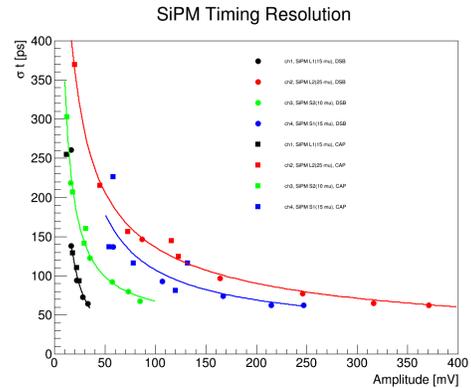

Figure 7. Energy resolution as a function of beam energy for the 4x4 W/LYSO array using capillary WLS readout out (red) compared to Y11 double clad WLS fiber readout (black), as measured using an electron beam in the CERN H4 beam line. These results indicate performance leading to a 1% constant term.

Figure 8. A test of a single LYSO/W module at the Fermilab Test Beam Facility FTBF. Single channel time resolution is shown with SiPM readout. Wave shifter readout was either DSB1 WLS dye in a multiclad optical fiber (dots) or DSB1 WLS in a liquid-filled capillary (squares).

These measurements, and further innovations described in Sections 3 and 4 below, form the basis of the current direction of RADiCAL R&D, emphasizing both high resolution timing and energy and timing measurement of EM showers within the same compact modular structure [11-12].

**3. Ongoing R&D on component instrumentation for RADiCAL.** The RADiCAL R&D has been subdivided into several tasks, including: Scintillator R&D; Waveshifter and Optical Element R&D; and Photosensor R&D.

**3.A. Scintillation R&D:** This task involves the investigation and development of bright, fast and radiation hard LYSO crystals, LuAG:Ce ceramics and $BaF_2$ crystals. Work continues with vendors to refine and test the quality improvements in LYSO:Ce crystals, LuAG:Ce,Ca ceramics and LuAG:Pr ceramics, and $BaF_2$:Y crystals. Irradiation testing is key to qualify advances in material production and quality.

- **Refinement of LYSO:Ce scintillators:**

Emphasis has been on refinement of crystal growth for purity, clarity, efficiency and radiation hardness. Recently, effort has been directed toward quality control and assurance for LYSO crystal productuon for the CMS BTL project, designed to measure the timing to ~30ps for minimum ionizing particles. Figure 5 shows the γ-ray induced photocurrent as a function of ionization dose rate applied to four BTL LYSO+SiPM packages [13]. Also shown in that figure are the numerical values of the F factor, defined as the radiation induced photo-electron number per second normalized to γ-ray dose rate ($F_γ$) or neutron flux ($F_n$), and the γ-ray induced readout noise (RIN:γ, $σ_γ$) of less than 35 keV, which is less than 1% of the 4.2 MeV MIP signal. The corresponding neutron induced readout noise (RIN:n, $σ_n$) is about 7 keV, more than a factor of four smaller than RIN:γ [13]. Figures 6 and 7 show the radiation induced absorption coefficient (RIAC) values as a function of proton [14] and 1 MeV equivalent neutron fluences [15] for LYSO/LFS crystals irradiated up to $10^{16}$/cm$^2$ by protons of 800 MeV in the blue room of the Neutron Research facility of the Los Alamos Neutron Science Center (LANSCE) and 24 GeV at the IRRAD facility at CERN, and neutrons in the east port of LANSCE, respectively. Consistent radiation induced absorption coefficient (RIAC) values at 430 nm for doses of $1.3 \times 10^{-14}$ $F_p$ and $1.4 \times 10^{-15}$ $F_n$ are observed, indicating a RIAC value of 0.5 m$^{-1}$ after the maximum fluences expected by the BTL LYSO crystals at the HL-LHC: $3 \times 10^{13}$ p/cm$^2$ and $3 \times 10^{14}$ $n_{eq}$/cm$^2$. Work along this line is onging, to investigate LYSO:Ce crystals at low temperature (-45°C) and radiation damage induced by ionization dose, protons and neutrons for the BTL detector at the HL-LHC, and for possible applications for the proposed FCC-hh.

- **Development of bright and fast LuAG:Ce and LuAG:Pr Ceramic Scintillators**

as efficient, cost-effective alternatives to LYSO. This includes emphasis on production of efficient ceramic materials, characterized for radiation hardness and scintillation efficiency. An interesting observation from this investigation is that the excitation wavelength of LuAG:Ce matches well the emission wavelength of LYSO:Ce [16], allowing LuAG:Ce crystal/ceramics to serve as an effective wavelength shifter for LYSO:Ce. This offers the prospect of all-inorganic ultra-compact, radiation hard, fast-timing EM calorimetry, and is currently under study by the group. LuAG:Pr has shown brighter and faster emission than LuAG:Ce, but initial measurements have shown it to be less radiation hard. Nevertheless, this material may still prove invaluable for its fast-timing potential at more central eta values: $η < 2.5$.

- **γ-ray, Proton and Neutron Induced Damage in bright and fast LuAG:Ce Ceramics.**

We have investigated the radiation hardness of LuAG:Ce scintillating ceramics. Because of its bright and fast fluorescence, excellent radiation hardness and potential low manufacturing cost, LuAG:Ce ceramics have attracted a broad interest in the HEP community and industry. Figs. 8-10 show radiation damage in LuAG:Ce ceramic samples irradiated up 220 Mrad at the JPL Total Ionization Dose facility [14], $1.2×10^{15}$ p/cm$^2$ at the PS-IRRAD facility of CERN and $6.7×10^{15}$ $n_{eq}$/cm$^2$ at the east port of LANSCE [15], respectively, and compare to LYSO crystals. Among the results: The RIAC values of LuAG ceramics against both neutrons and protons are approximately a factor of two smaller than that of LYSO crystals.

With 90% light output remaining in 1 mm thick samples after neutron and proton fluences of up to $6.7\times10^{15}$ $n_{eq}/cm^2$ and $1.2\times10^{15}$ $p/cm^2$ respectively [14,15]. Thus, LuAG ceramics are a promising inorganic scintillator for future HEP experiments in a severe radiation environment, such as the HL-LHC and the proposed FCC-hh. R&D along this line will continue to investigate co-doped LuAG:Ce,Ca ceramics from Radiation Monitoring Devices, Inc. supported by a DOE SBIR award [17] to further improve optical quality, the Fast/Total ratio of fluorescence decay and radiation hardness of the LuAG ceramic materials.

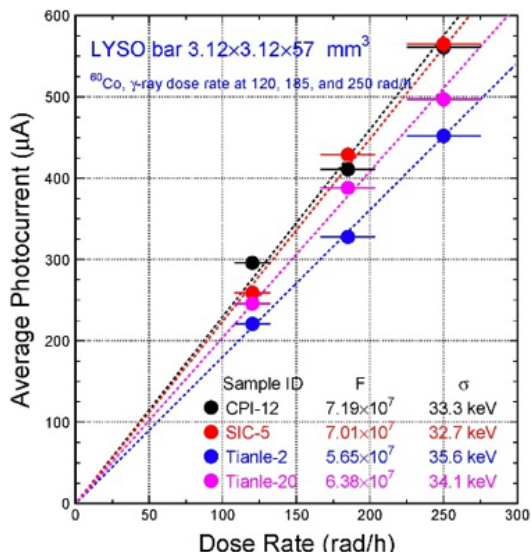

Figure 9. γ-ray induced photo current as a function of ionization dose rate up to 250 rad/h for four LYSO bars from 3 vendors [13].

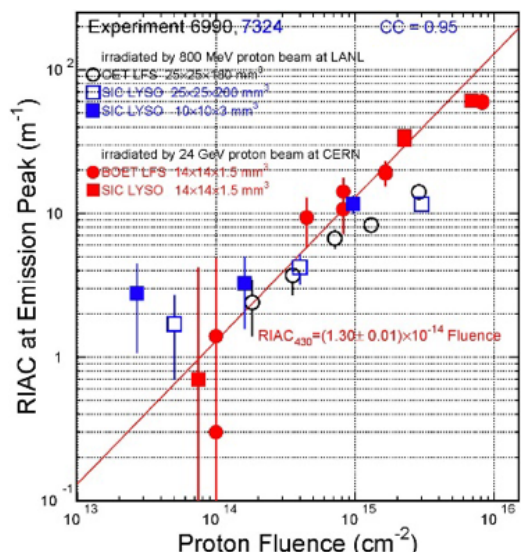

Figure 10. RIAC values as a function of the proton fluence are shown for LYSO/LFS crystals irradiated at CERN and LANSCE [14].

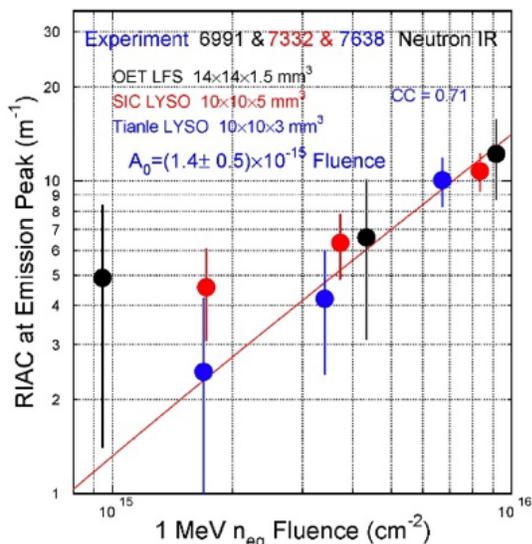

Figure 11. RIAC values as a function of the 1 MeV equivalent neutron fluence are shown for LYSO/LFS crystals irradiated at LANSCE [15].

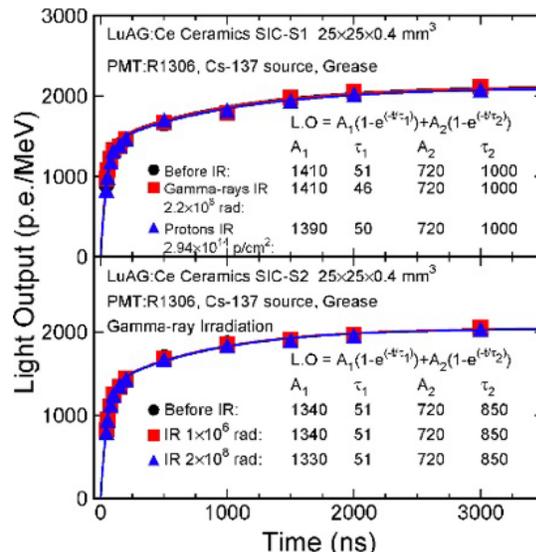

Figure 12. Light output as a function of integration time is shown for LuAG:Ce ceramics after 200 Mrad γ-ray and $3 \times 10^{14}$ $p/cm^2$ fluence [16].

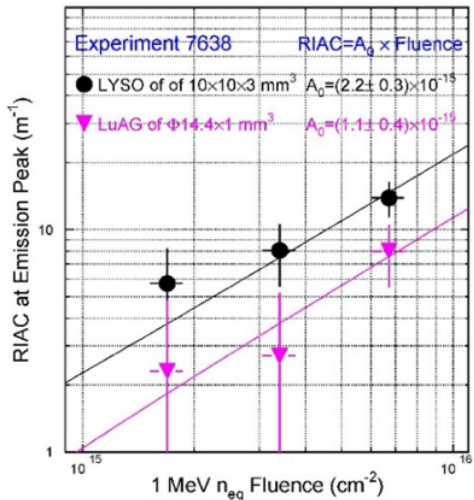 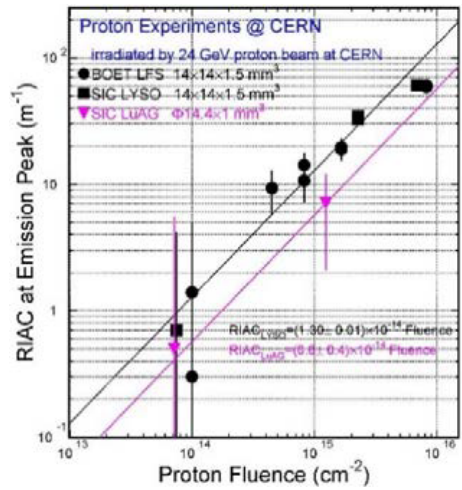

Figure 13. RIAC values as a function of proton fluence for LYSO/LFS crystals and LuAG ceramics irradiated at CERN [14].

Figure 14. RIAC values as a function of 1 MeV equivalent neutron fluence for LYSO crystals and LuAG ceramics irradiated at LANSCE [15].

- **BaF$_2$:Y Scintillator for Fast Timing Applications.**

Yttrium doped Barium Fluoride (BaF$_2$:Y) is under consideration for Mu2e-II ultrafast calorimetry and could also find application in hadron collider environments. The yttrium suppresses significantly the slow component of the BaF$_2$ emission at $\lambda$ = 300 nm with 600 ns decay time, while maintaining unchanged the ultrafast component at $\lambda$ = 220 nm with less than 0.6 ns decay time. Figure 15 shows X-ray excited emission (XEL) spectra for BaF$_2$ with different yttrium doping levels, showing that the slow 300 nm component can be reduced with little effect on the fast 220 nm component [18-21]. Figures 16 and 17 show proton [14] and neutron [15] induced radiation damage in BaF$_2$ crystals is larger than LYSO at a low fluence, and is comparable to LYSO at a high fluence, which is consistent with ionization dose induced damage in large size (20 cm long) BaF$_2$ crystals [21]. Because of its high light yield in the 1$^{st}$ ns of excitation and suppressed slow component, BaF$_2$:Y crystals are also considered for precision timing applications, and for GHz hard X-ray imaging for a future Free Electron Laser Facility such as DMMSC [22-24], and further investigation is warranted for HL-LHC and FCC-hh applications.

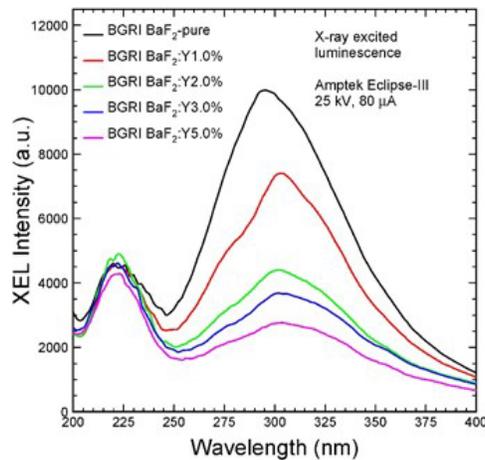

Figure 15. XEL spectra of BaF$_2$ with different yttrium doping level, showing the slow component suppression with little effect on the fast component [19].

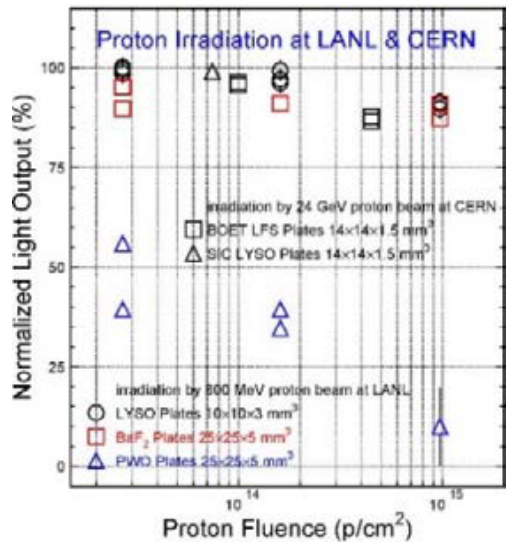 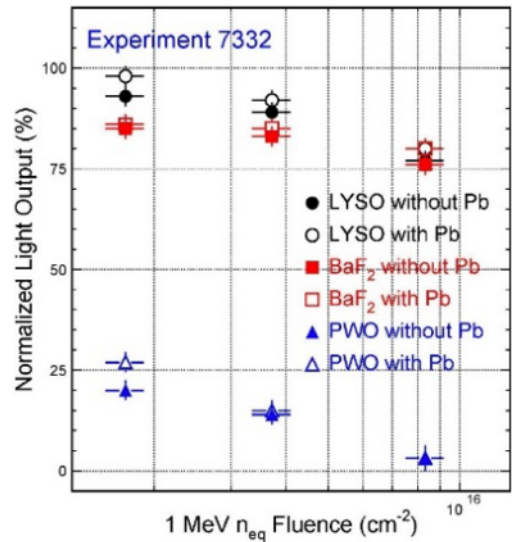

Figure 16. Normalized light output as a function of the proton fluence is shown for LYSO, BaF2 and PWO plates irradiated at CERN and LANSCE [14].

Figure 17. Normalized light output as a function of 1 MeV equivalent neutron fluence is shown for LYSO, BaF2 and PWO plates irradiated at LANSCE [15].

- **Scintillator R&D Summary**

Baseline: The current baseline material for RADiCAL is LYSO:Ce crystal, which has an emission wavelength $\lambda$ = 425nm that is well matched to fast wavelength shifters.

R&D Goals: We have found that the radiation hardness of the scintillating ceramic LuAG:Ce is better than LYSO crystal (factor ~2), indicating promise for FCC-hh applications. With an emission wavelength $\lambda$ = 520nm, an appropriately matched new waveshifter for light collection is required. In this case, the LuAG:Ce ceramic plates would replace the LYSO:Ce plates in the module shown in Fig.1.

Further important options under consideration for scintillation plates include: LuAG:Pr (emission $\lambda$ ~ 310nm, for fast timing options), $CeF_3$ (emission $\lambda$ ~ 330nm, for radiation resistant options), and $BaF_2$:Y (emission $\lambda$ ~ 220nm, for ultrafast timing options). Such plates could be positioned strategically in the shower max region for fast timing detection and position measurement. See Table 1 below for summary.

**3.B. Waveshifter (WLS) R&D:** WLS capillaries and fiberoptic filaments form the optical "bridges" that connect the scintillation light emission from the tiles within RADiCAL modular structure to photosensors positioned at upstream and downstream ends of a module. The choice of the WLS material depends upon wavelength of the scintillation emission and the wavelength response of the photosensors used.

- **Waveshifters Matched to LYSO:Ce Emission:**

There are several WLS candidates of interest to shift the $\lambda$ ~ 420nm emission of LYSO:Ce. The first is DSB1, developed by Notre Dame in collaboration with Eljen Technology, Sweetwater, TX, having a factor of 2.5x faster decay time than J2/Y11 while maintaining similar fluorescence efficiency. With emission at $\lambda$ ~ 495nm, DSB1 is well matched to SiPM photosensors. A second WLS candidate is ceramic LuAG:Ce, measured to be more rad hard than LYSO:Ce (see Figures 9 and 10 above).

- **Waveshifters Matched to LuAG:Ce Emission:**

Quantum Dot wavelength shifters in glass and/or polysiloxane materials are to be developed, to waveshift the long-wave ($\lambda$ = 520nm) emission of LuAG:Ce into the spectral region 540< $\lambda$ < 570nm where SiPM photosensors remain efficient. In these investigations, Quantum Dot composition, size, solubility and material host composition are all considerations.

- **Waveshifters Matched to LuAG:Pr emission:**

There are several candidates of interest here to shift the $\lambda$ ~ 310nm UV emission of LuAG:Pr. Appropriate for fast timing and used in wave shifting elements of short length is the dye p-Terphenyl (pTP) which is fast, highly efficient, and soluble in organic plastics and liquids. For energy measurement, dyes based on the flavenol 3-hydroxyflavone [25,26] are under development with Eljen Technology. The best of these, Flavenol-560, can shift UV emission to $\lambda$ ~ 560nm with good efficiency and decay time ($\tau$ < 4ns) [27,28].

- **Detection of BaF$_2$:Y emission:**

Because of the challenging VUV emission of BaF$_2$:Y and the necessity to avoid the predominant slow component of the emission peaked at 300nm, specialized photosensor development (diamond photosensors) are under consideration to pursue this option further, rather than using waveshifting techniques. This R&D avenue is being explored in collaboration with Physical Sciences Inc, Andover, MA.

## 3.C. Optical Transmission Elements for Energy and Timing Measurement.

These elements can include WLS liquid-filled quartz capillaries or WLS ceramic filaments. Recent developments have been focused on the improvement of both light collection and transmission to the benefit of both energy and timing resolution [29-31]. The characteristics of several capillary versions are shown in Figures 18-24.

- **Capillaries for energy measurement (E-Type Capillaries and Filaments).**

The WLS liquid-filled capillaries have a short ruby quartz-filament at the readout end for core blocking of light trapped within the core liquid. This serves to minimize the effects of non-uniformity in light collection introduced by the optical attenuation of light propagating through only the core liquid in the capillaries. That this has been successfully achieved is shown in Figure 16, where the WLS light collection is uniformly flat as a function of length, and behavior that is maintained with increasing ionization dose up to 150Mrad. This dose level corresponds to 3000fb$^{-1}$ of operation in a hadron collider endcap region near $|\eta|$ ~ 3, and obviates concerns about effects of differential longitudinal non-uniformity to the constant term in the energy resolution. However, should additional light-collection be needed, for example to transmit the light via fiber-optic waveguides to remotely placed, radiation-sensitive photosensors, clear-ended capillaries (Figures 18,20,21) would provide an acceptable solution.

In irradiation studies, we observe in all versions of the full length DSB1 WLS liquid-filled capillaries (ruby core blocked or clear ended) a ~11% reduction in detected light transmission for every 50 Mrad of gamma exposure. After three such exposures, corresponding to 150Mrad of total dose, one finds ~70% of light transmission remaining as shown in Figures 19-21 [30,31]. Similar results have been found during exposure of the capillaries to 800 MeV protons at LANL up to ~ 10$^{15}$ p/cm$^2$, consistent with the endcap environment near $|\eta|$ ~3. After this exposure, again 70% of light was found to be remaining. These data indicate that the sealed WLS liquid-filled capillaries have the potential to be performant structures for use in future collider applications.

A new development being explored is the incorporation of inorganic waveshifters within the, capillary cores or as stand along filaments, avoiding the usage of liquid-filled capillaries altogether. This is accomplished through the use of full-length LuAG:Ce ceramic filaments inserted into RADiCAL modular structure, replacing the DSB1 liquid-filled capillaries.

- **E-type capillaries/filaments Summary:**

Baseline: Sealed quartz capillaries containing EJ309 liquid and DSB1 waveshifter with emission at $\lambda$ = 495nm.] These are well matched to both the LYSO:Ce scintillator and to conventional SiPM readout as well as GaInP or more radiation hard alternatives, once developed.

R&D Goals: LuAG:Ce ceramic filaments can serve as waveshifter for LYSO:Ce, with emission at $\lambda$ = 520nm, well matched to SiPM, GaInP and new devices if available. To waveshift the scintillation light from LuAG:Ce ceramic plates, LuAG:Pr ceramic plates or $CeF_3$ crystal plates, different WLS materials are needed. Candidates include flavenols [28], pTP and TPBD shifters among others in organic plastic and siloxane filaments. See Table 1 below for a summary.

- **Capillaries for Timing Measurement. (T-type Capillaries and Filaments)**

We are developing specialized capillary structures to read out the WLS light from the region of EM Shower Max in RADiCAL modules. The intent is a high-resolution, precision-timing measurement, to complement the energy measurement. As shown in the Figure 22, a WLS filament is embedded in a quartz capillary at the corresponding location of shower max in a RADiCAL modular structure. Figure 23 shows such a specialized capillary with a DSB1 organic plastic filament in comparison to two other DSB1 iquid-filled capillaries used principally for energy measurement. The localized filament for shower max timing is clearly visible. Figure 24 shows the detected WLS signal from a timing capillary, excited along its length by a UV LED at 420nm, in comparison with that of a liquid WLS-filled capillary for energy measurement, indicating the very strong response from the region of shower max and negligible response elsewhere.

Once the shower max WLS filament is inserted, the remaining core volume of the capillary is filled with clear rad-hard quartz fiber of appropriate diameter and length, inserted from upstream and downstream ends, which completely fill the full length of capillary core. These are then fused and sealed within the capillary, creating a solid rad-hard quartz optical fiber upstream and downstream of the WLS element. This arrangement provides transmission of the WLS light with minimal optical absorption to photosensors. In this type of capillary structure, the WLS filament acts effectively only as a light source and has minimal impact on the optical transmission of the WLS light within the remainder of the capillary structure. The key parameter is the stability of the WLS material itself under irradiation, rather than light transmission through WLS material, as the volume of shifter is a very small fraction of the capillary length. The timing of the light signal can be detected by photosensors positioned at both the upstream and downstream ends of the capillaries, and hence directly at the ends of the RADiCAL module. Interestingly, since the timing capillary is mostly rad-hard quartz, the downstream photosensor will detect WLS light and also detect Cerenkov light created in response to EM shower particles passing through the quartz. The upstream photosensor, located opposite to the direction of the shower particles, will detect primarily WLS light, which is isotropically produced.

Candidates for the WLS elements are 15mm-20mm long filaments inserted into the quartz capillary core to cover three to four radiation lengths of the RADiCAL sampling structure in the region of EM Shower Max. Filament materials have been selected for high efficiency, fast response and rise time, and potential for radiation hardness. These include DSB1 organic plastic filaments, LuAG:Ce ceramic filaments and polysiloxane, organic polymer and glass filaments containing Quantum Dots.

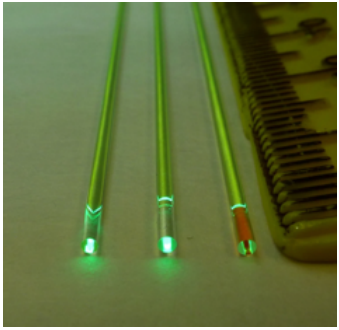

Figure 18. Capillaries under UV illumination with different end sealing: Right - ruby core blocked; Middle - clear core sealed; Left - clear end fused.

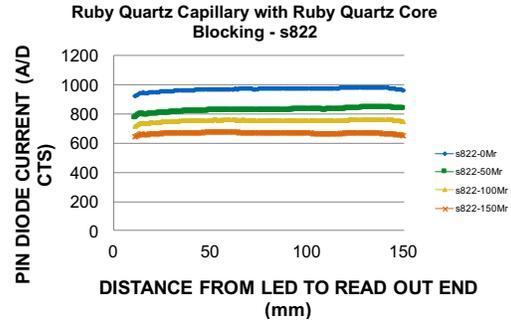

Figure 19. Light transmission as a function of distance to the readout end of a ruby core-blocked capillary from the WLS excitation location.

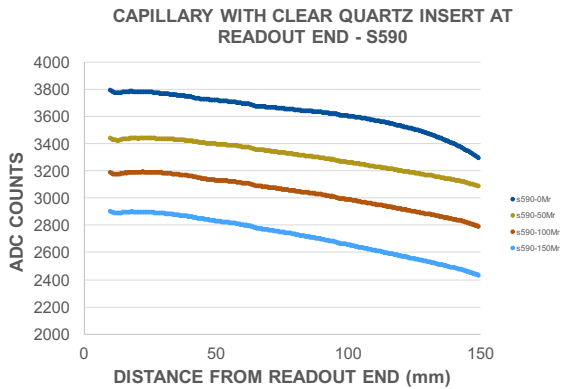

Figure 20. Light transmission as a function of distance to the readout end of a clear quartz insert capillary from the WLS excitation location.

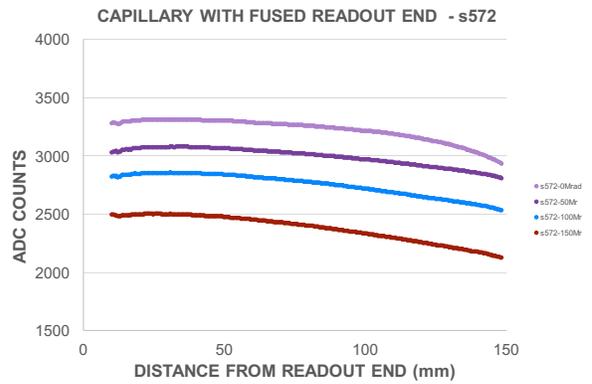

Figure 21. Light transmission as a function of distance to the readout end of a clear end-fused capillary from the WLS excitation location

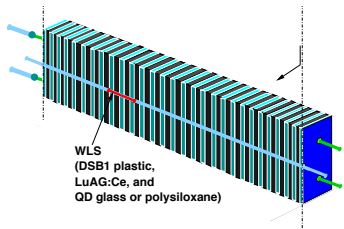

Figure 22. A shower max timing capillary with WLS filament (red) located at the shower max region. Incident electrons enter the left-hand end of the module.

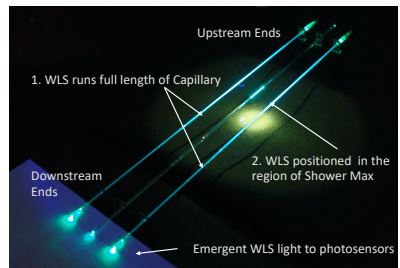

Figure 23. DSB1 WLS capillaries under UV illumination. Upper left and Lower right are WLS liquid filled capillaries primarily for energy measurement. Middle is a Shower Max timing capillary with DSB1 organic plastic filament located near shower max.

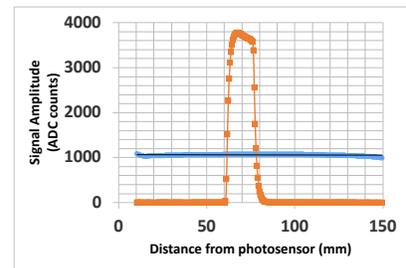

Figure 24. Light transmission as a function of distance from the WLS excitation location to the readout end of a ruby core-blocked energy capillary (blue line) with DBS1 WLS, compared to a timing capillary with DSB1 WLS filament (orange curve). Waveshifter was excited by an LED at 420nm.

- **T-type capillaries/filaments Summary**

Baseline: Quartz capillaries containing DSB1 organic plastic filaments, with emission at λ = 495nm and positioned near shower max. These are well matched to LYSO:Ce scintillator and SiPM and GaInP photosensors.

R&D Goals: LuAG:Ce ceramic filaments replace the DSB1 filaments at shower max and are optically coupled to photosensors via quartz rods. The LuAG:Ce material is optically well matched to SiPM, GaInP photosensors.

**3.D. Photosensor R&D:** This Task has two elements: The development of SiPM for improved performance in efficiency, timing, signal recovery, and operation in elevated levels of radiation; [32-33] and a "blue-sky" and hence longer time-frame effort to develop pixelated, Geiger-mode photosensors based upon the large band gap material GaInP [34] and others.

- **SiPM Photosensor Development:**

We have been working with with two vendors – Hamamatsu (HPK) in Japan and FBK in Italy. SiPM prototypes have been fabricated at both companies, with subsequent bench and radiation testing. Both vendors are working to improve the "temperature coefficient" of the SiPMs, and the RADiCAL group has been instrumental developing onboard thermoelectric cooling (TEC). The TEC is central to the operation and performance of the CMS Barrel Timing Layer (BTL).

The group has been developing small-pixel SiPM, whose ultimate radiation tolerance would be sufficient for applications where the radiation levels are more moderate and/or where access is possible for device replacement. The emphasis on small pixels is to address the issues of light saturation, rise time and fast recovery in silicon-based technology. Progress has been made through two generational steps with FBK in producing devices with low gain, good PDE, and small pixels of 5µm, 7µm and 10µm, with refined field shaping and novel pixel architecture. [35,36] Figure 25 displays pixel layouts for two sizes of devices, and Figure 26 shows a pulse image from the first fabrication of 5µm pixel devices and indicates fast response (full pulse rise in ~ 3ns) and rapid recovery (pulse decay in ~8ns). Figure 27 displays the photodetection efficiency (PDE) for devices of different pixel sizes as a function of the overvoltage V-$V_B$ where $V_B$ is the breakdown voltage. It should be noted that even the lower PDE values in Figure 27 are more than adequate for energy and timing measurements with the photosensors proximately placed on the ends of a RADiCAL module, which is the approach now being followed.

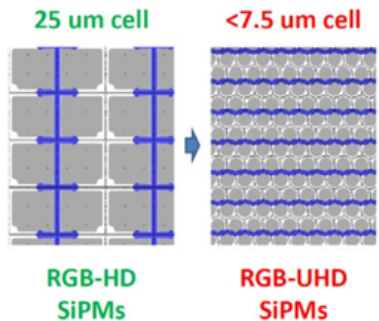

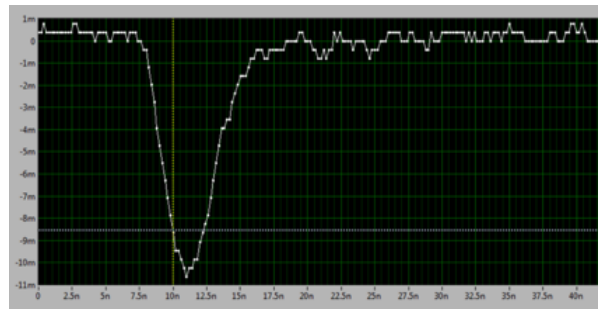

Figure 25. New SiPM Pixel Architectures from FBK.

Figure 26. Output pulse from a FBK SiPM, with 5µm pixel size, under laser excitation. Horizontal axis has divisions every 2.5ns; vertical axis in mV units.

Hamamatsu (HPK) devices with pixels of 12µm have been shown to operate successfully in fluences up to 2.2 x $10^{14}$ p/cm$^2$ when operated at a temperature T = -20°C and when exposed to reactor neutrons up to 2 x $10^{13}$ n/cm$^2$. At the present level of performance, these SiPM could be used effectively when positioned remotely from a W/LYSO:Ce RADiCAL module, or could be attached directly to scintillators in locations where the radiation dose is more moderate, for example deeper in endcap calorimetry, or in the central region of collider detectors. With further R&D on SiPM device development and cooling, the operational range of these devices might be extended to > $10^{14}$ n/cm$^2$ or more depending upon the application. Such studies are currently in progress and under development by CMS for the Barrel Timing Layer (BTL).

The TEC, placed locally and proximately to the SiPM, can be operated as a "heat pump". In simulations by CMS and in consort with the CMS $CO_2$ cooling system: in forward bias, the TEC can readily cool the SiPM to temperatures -30c ≤ T ≤ -45c during data taking; in reverse bias, the TEC can raise the temperature of the SiPM significantly (up to +40c) so that the devices can be annealed during intervals of no beam or shutdowns (Fig.28). This strategy can inform further developments in SiPM operation in moderately elevated radiation fields [37].

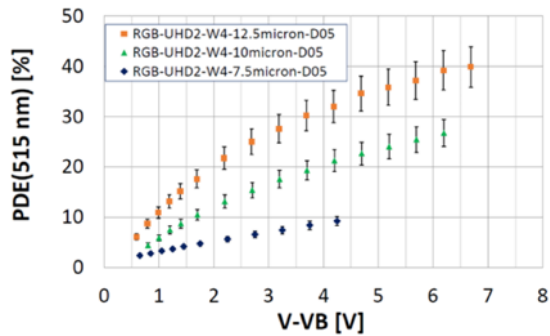

Figure 27. Photodetection Efficiencies as a function of bias voltage for SiPM from FBK, with pixels of 5um (black), 7um (green) and 10um (orange).

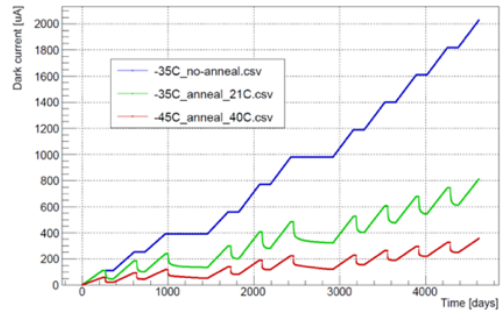

Figure 28. SiPM Dark current over time for different annealing scenarios for the CMS BTL utilizing HPK SiPM. Blue curve – no annealing. Green and Red curves - both annealing and cooled operation.

- **Large Bandgap Photosensor Development.**

In collaboration with Lightspin Technologies Inc, GaInP photosensors are underdevelopment, whose large band gap offers the potential for significant improvement in radiation tolerance over conventional Silicon-based (SiPM) devices and possible room temperature operation [36]. To date, the development has proceeded through an informative process of five generational steps. The 5$^{th}$ generation of devices (gen5.1) with 10µm and 25µm pixels (Fig.29), have shown single-photon counting capability when operated in Geiger-mode in a manner similar to SiPM (Figs.30-32). These devices have AlGaAs windows and anti-reflective coatings, and some variants had bypass capacitors to enhance speed and amplitude of the signal pulses. Bench tests and irradiation studies with these devices have shown promise, but the studies also indicate that more work is needed on surface preparation. The fundamental premise that bulk material properties contribute little to the increase of dark noise appears to be validated.

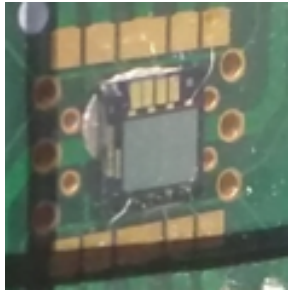
Figure 29.  Photo of a 4x4 mm$^2$ GaInP Photosensor consisting of 10 arrays, 0.5 x Will c1.5mm$^2$ size and 25 μm pixels

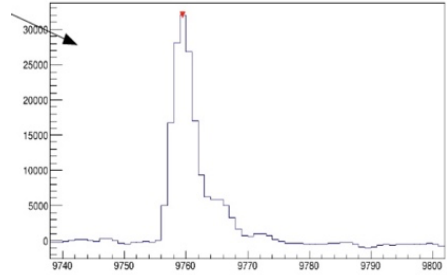
Figure 30.  Signal Pulse from a GaInP 25μm SPAD array under laser excitation.

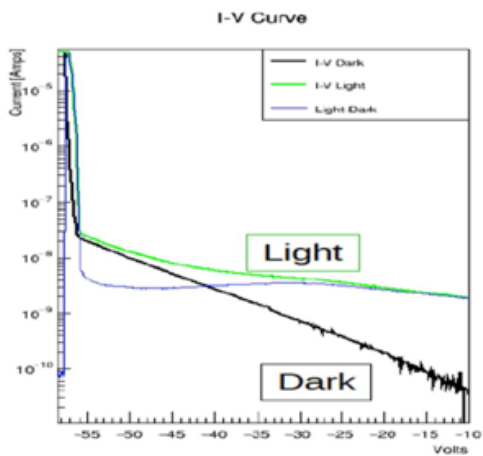
Figure 31.  IV curves for GaInP photosensors under illumination (green, red curves) and dark field (blue). Horizontal axis is bias voltage. Vertical axis is output current.

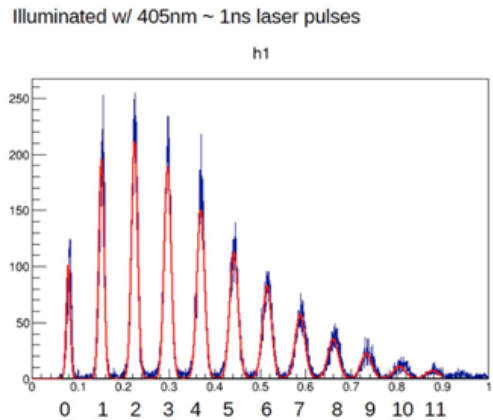
Figure 32.  GaInP Photospectrum shows individual photopeaks. Left most (0) is the pedestal.   Illumination of the device at λ=405nm.

A revision of 5th generation GaInP devices (gen5.2) was initiated with a different foundry, to mitigate surface dark currents by varied surface treatments.  Changes included an Al-ln-P window, improved fill factor and improved anti-reflective coating.   However, these devices were found to be noisier than version 5.1.  A 6$^{th}$ generation version was initiated to compare the performance of an alternate epitaxy approach, as well as development of devices with smaller pixels (5-7 μm).  But this version suffered from significant surface currents.

Given the current status of these developments, it was determined that there was need to engage a broader range of commercial vendor expertise. For this, planning is underway by the RADiCAL group to explore further the large band gap photosenseor parameter space including GaInP, SiC, diamond photosensors and other materials, to cover a broad spectral range from 200nm < λ < 650nm, to provide compatibility with various scintillator and wavelength shifter options, in combination with improved radiation tolerance.

- **Photosensors R&D Summary**

Proximate placement of photosensors directly on the RADiCAL modules directly benefits timing performance, and development of rad hard sensors is a priority of the R&D effort.

Baseline: SiPM are the baseline choice for establishing what is possible for energy, timing and position measurement in our current testing. Devices from several vendors, HPK and FBK, have been utilized. However, the radiation hardness of the currently developed SiPM devices is insufficient for their proximate positioning on the modules under FCC-hh conditions [32,33]; instead, their use would require positioning remote from modules with optical interconnection via quartz waveguides as indicated in Fig.1.

R&D goals: For proximate positioning, rad hard alternatives to SiPM are needed and large bandgap pixelated Geiger-mode devices and APDs are an important developmental direction. Material/device choices are driven by the wavelength range needed for light detection.
- For visible light detection in the 400nm < $\lambda$ < 600nm range, GaInP and novel Visible Light Photosensors are under consideration with commercial partner, Physical Sciences Inc.
- For UV light detection in the 300nm < $\lambda$ < 400nm range, SiC photosensors are an appropriate choice and are under consideration for direct detection of light emission from LuAG:Pr and $CeF_3$.
- For VUV application, for example direct detection of light emission from $BaF_2$:Y (200nm < $\lambda$ < 220nm), diamond photosensors are the appropriate choice. Planning for development of new photosensors across this broad spectral range is underway with a commercial partner, Physical Sciences Inc.

If sufficiently rad hard, photosensors could be mounted within the interior of a RADiCAL module and optically coupled directly to scintillation plates positioned at shower max for precision timing measurement. This would avoid the use of WLS materials altogether for timing. Initial tests of this concept are planned using a single Radical module instrumented with four SiPM positioned on a LYSO:Ce crystal tile near shower max to measure the timing and position response of this specialized configuration. Table 1 summarizes the various material and photosensor combinations under investigation for RADiCAL, covering the wavelength range from 200nm < $\lambda$ < 600nm.

**Table 1. RADiCAL R&D: Radiation Hard Scintillator, Wavelength Shifter and Possible Photosensors**

| Scintillator material | Scintillator Emission Wavelength | Wavelength Shifters | WLS Emission Wavelengths | Photosensor Possibilities |
|---|---|---|---|---|
| LYSO:Ce | 425nm | DSB1 | 495nm | SiPM, GaInP, new |
| LYSO:Ce | 425nm | LuAG:Ce | 520nm | SiPM, GaInP, new |
| LYSO:Ce | 425nm | Direct - No WLS | | SiPM, new |
| LuAG:Ce | 520nm | Quantum Dots | 560-580nm | SiPM, GaInP, new |
| LuAG:Pr | 310nm | pTP / TPB / Flavenols | 360nm / 460nm / 530-560nm | SiPM, GaInP, new |
| LuAG:Pr | 310nm | Direct - No WLS | | SiC |
| $CeF_3$ | 330mm | pTP / TPB / Flavenols | 360nm / 460nm / 560nm | SiPM, GaInP, new |
| $CeF_3$ | 330nm | Direct - No WLS | | SiC |
| $BaF_2$:Y | 220nm | Direct – No WLS | | Diamond |

**4. Beam Testing Plans**: Underway are EM Energy and Timing Measurements using an individual RADiCAL module, to be followed by development and study of arrays of such modules to provide full EM shower containment. The primary emphasis is to test, with electron and hadron beams, RADiCAL modules with successively more radiation hard components (scintillators, wave shifters and optical transmission elements), for energy resolution, timing resolution and shower pattern analysis, utilizing the new materials developed by the group and others as described in Section 3 above. The photosensors to be used initially are SiPM as they are readily available for the studies, but could be replaced by more rad hard devices was they become available.

**4.A Simulations informing the research planning:** GEANT4 simulations have been carried out for a 3x3 array of RADiCAL modules of the type shown in Fig. 1 and displayed schematically in Fig. 33, and qualified from early measurements described in Sect. 2 above [38]. The cross-sectional area of each module is 14mm x 14mm, consistent with the Molière Radius.

For the energy measurements, the signals are obtained from the energy capillaries which run the full length of each module (covering 25 $X_o$). For electrons incident near the center of the central module of the 3x3 array, ~80% of the shower energy is contained within the central module (Fig.34) and ~100% is contained within the totality of all nine modules (Fig.35). This characteristic behavior has been found to apply across a very wide range of energies as Figures 34-35 show. In a manner routinely used in HEP experiments, the fitting of the spatial profile of observed energy in the modules provides the spatial location of center of the EM shower and, in a test beam with tracking information, can be compared with the location of the incoming beam particle incident on the array.

The RADICAL modules also provide timing measurement, determined at shower max where the number of charged shower particles is greatest. GEANT4 simulation has been used to study the timing resolution derived from this region. For this study, the simulation assumed only one SiPM readout at the downstream end of a shower max WLS timing capillary in a single RADiCAL module (Fig.22), although in reality there will be timing measurements available from both upstream and downstream ends, and several such capillaries employed. But with only downstream readout, the GEANT4 simulation indicated excellent timing resolution can be achieved dependent upon detected light level [39]. Noteworthy is that the shower max timing signal is derived from a region of very small transverse size r < 3mm (Fig.36), a region whose radius is significantly smaller than the Molière Radius. Also noteworthy is that this small region contains one to two orders of magnitude more charged shower particles than would mip signals passing simultaneously through the scintillation layers at that location, creating a very large and localized optical pulse. The timing resolution will be dominated by the measured rise time of this signal and will depend also on the detected light yield within the first nanosecond of the optical pulse. Figure 37 indicates that, in simulation, timing resolutions of < 50ps could be achievable with RADiCAL modules. By reading out the light from both upstream and downstream ends of timing capillaries, and by using several timing capillaries per module, further improvement in the timing resolution measurement would be anticipated.

The shower max method also provides an additional pattern recognition constraint. In the modular array, the module that is struck by the incoming electron (or gamma) provides the predominant timing signal for the EM shower. In adjacent modules, the timing signals are effectively absent, since there is effectively no energy deposited in them in that region, given the very compact transverse size of the shower at the shower max location. This is potentially powerful locational tool to identify which module is struck by an incoming EM particle in the more forward (higher $\eta$ regions) of collider detectors where separation and reconstruction of overlapping showers will be challenging.

These simulations suggest that the shower max timing technique, combined with fast, ultracompact energy measurements are together a powerful technique for measuring: (1) EM energy, (2) timing and (3) pattern determination. These capabilities have motivated tests of modular arrays of RADiCAL modules, in which the behavior of the array will be assessed with beam particles striking within modules, near modular boundaries, and under conditions when multiple beam particles are present.

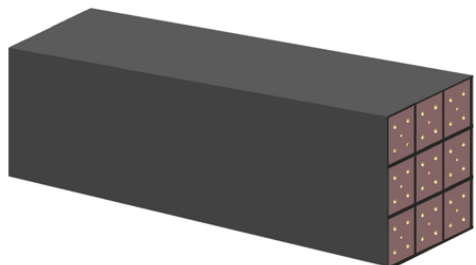

Figure 33. RADiCAL Modular Array for GEANT4 Shower Energy Containment Studies, here shown as a 3x3 array of modules of the type showm in FIGURE 1. Beam electrons enter through the left-hand end at normal incidence.

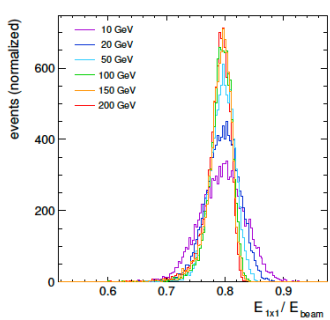

Figure 34. Energy contained in the "central" module of a 3x3 square array, where the incoming electron hits the center of the array. GEANT4 simulation.

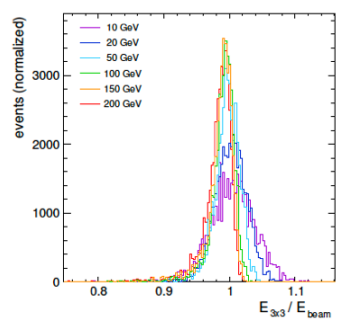

Figure 35. The total energy of an incident electron contained within all nine modules of a 3x3 square array of modules. GEANT4 simulation.

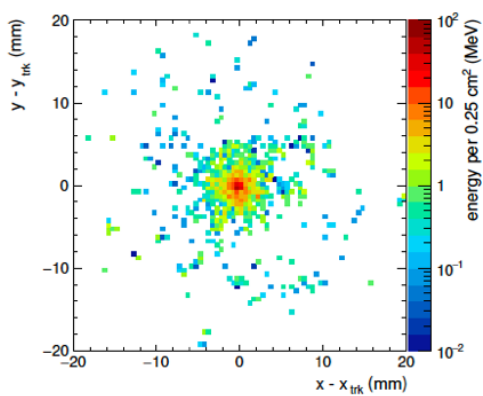

Figure 36. Location of the energy in a RADiCAL module at shower max for 50GeV electrons in GEANT4 simulation. The boundary of the module itself is a square 14mm on a side located at the center of the plot.

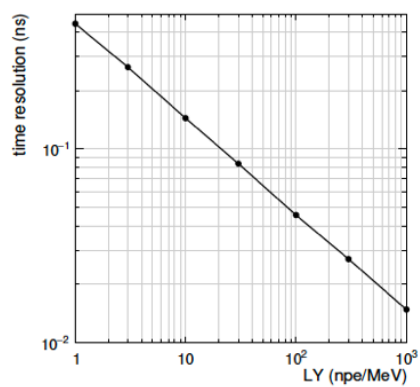

Figure 37. Timing resolution vs detected light yield in photoelectrons per MeV, simulated for a 50GeV electron shower. Downstream readout only in this study. GEANT4 simulation.

**4.B. Modules and modular arrays for beam testing.** Figure 38 shows a schematic of the upstream face of two types of RADiCAL modules, as the beam would see them as it enters the modules. Indicated are the transverse placement locations of four or six WLS capillaries/filaments which can be configured in different ways to measure energy, time and position. Figures 39 and 40 show test arrays of such modules, of transverse size sufficient to contain the full shower energy of an incoming beam electron.

- E-type capillaries, primarily for energy measurement:
    - For our initial goals, the E-type capillaries are of 1.0mm outer diameter and 0.4mm inner diameter, with their cores filled with EJ309/DSB1 liquid wavelength shifter over their full length and are read out with SiPM and low-gain amplifiers at the downstream ends only.
    - For our primary and stretch goals, rad hard LuAG:Ce WLS filaments of 114mm length and 1.15mm diameter will replace the liquid-filled capillaries with two benefits: photosensors can be placed at both upstream and downstream ends to measure the signal energy and timing; and the use of liquid-filling is avoided. This will be practical ultimately, if photosensors with sufficient radiation hardness are developed and available.
- T-type capillaries, primarily for timing measurement:
    - For our initial goals, the T-type capillaries are of 1.15mm outer diameter and 0.95mm inner diameter with 15-20mm long DSB1 organic plastic WLS filaments positioned in the cores at shower max. The remainder of the capillary cores upstream and downstream of the WLS filaments are filled with quartz rods and fused to the capillary walls forming solid quartz waveguides.
    - For our primary and stretch goals, rad hard LuAG:Ce WLS filaments of 1.0mm diameter and 15-20mm length will be positioned near shower max. These filaments will be optically connected via quartz rods of 1.0mm diameter to photosensors positioned at upstream and downstream ends.

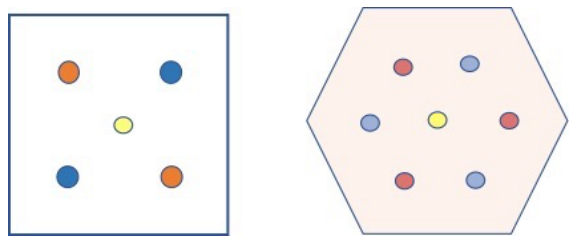

**Figure 38.** Schematics of capillary placement in RADiCAL Modules of (left) square cross section and (right) hexagonal cross section. The distance between opposite parallel sides is set by the Molière radius (14mm). The red circles correspond to the position of E-Type capillaries for energy measurement, the blue circles correspond to the position of T-Type capillaries for timing measurement. The yellow central fiber is for module calibration through injection of laser signals via "leaky" optical fiber. The hexagonal module benefits from an extra E-Type and T-Type capillary providing improved spatial and timing precision.

For shower position determination, the spatial localization of an EM shower is provided by the signal amplitudes of the energy measurements from the capillaries. Note that at shower max, the shower radius is significantly smaller than the Molière radius (< 5mm vs ~13.7mm) [5]. Capitalizing on this, the shower position can be localized within a module to within a few mm, beneficial for event reconstruction under high pileup conditions in endcap and forward regions of experiments and for distinguishing nearby showers from decays of highly-boosted objects.

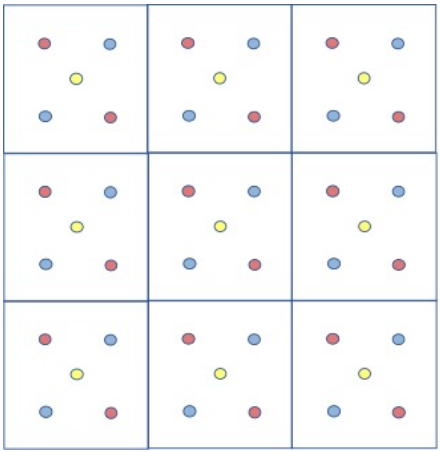 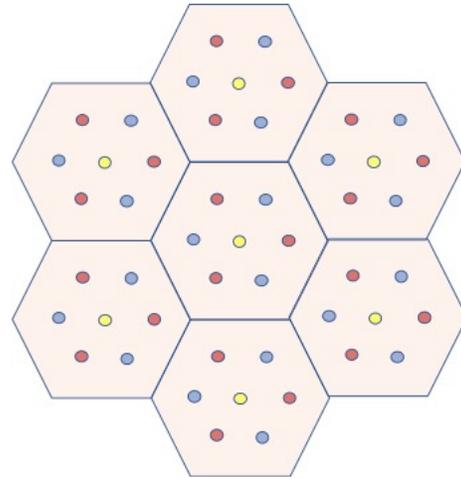

Figure 39. Beam view of a 3x3 modular array of RADiCAL Modules of square cross section, which consists of a central module surrounded by eight others. In each module: two red dots are E-Type capillaries; two blue dots are T-type capillaries; one yellow dot is a calibration fiber.

Figure 40. Beam view of a seven-module array of RADiCAL Modules of hexagonal cross section, which consists of a central module surrounded by six others. In each module: three red dots are E-Type capillaries; three blue dots are T-type capillaries; one yellow dot is a calibration fiber.

In late December 2021 at the Fermilab Test Beam Facility (FTBF), a test was carried out of a single RADiCAL module of square cross section instrumented with two E-type and two T-type capillaries, using negative beams of energy 12GeV < E < 28GeV as well as protons; these data are currently being analyzed. Future tests are planned at the Fermilab FTBF (beam energies up to 28 GeV) and at CERN H8 (beam energies up to 100 GeV). For the higher energy tests, modules of both square and hexagonal are under consideration. GEANT4 simulation will help guide the fiber positioning in the two options.

**5. Summary and Conclusions**

The pattern recognition power and the potential for high resolution measurement of both timing and energy for EM objects (electrons, positrons and gammas) in testing of arrays of RADiCAL modules (square and hexagonal cross sections) is compelling. The extensive R&D by the RADiCAL group on the components that comprise the modules, and the measured extensive radiation hardness of these elements, gives confidence that these structures can provide a promising technique for EM Calorimetry in future high luminosity hadron collider experiments.

The primary objective of the R&D is to build modular arrays and test them for energy reolution and precision timing with the various radiation hard technology options that have been and continue to be developed by the group. A series of beam tests is planned, using materials of successively higher radiation tolerance, with the aim of attaining the primary goal as stated in the introduction, of having highly performant EM calorimetry to $|\eta| \leq 4$ in the hadron collider environment. Additionally, the tests are expected to provide an important step forward toward our stretch goal of reaching further into the forward direction ($\eta > 4$) with precision EM calorimetric techniques. And, these RADiCAL goals are directed toward addressing the Priority Research Directions (PRD) for calorimetry listed in the report of the DOE Basic Research Needs (BRN) workshop for HEP instrumentation.[1]

Lastly, the impact of this research and development program, while not experiment specific, is potentially broad and significant, and can be expected to inform further developments in high energy physics instrumentation. The technologies are versatile and can be applied in particle physics, nuclear physics, materials science, basic energy sciences and extended to medical physics and homeland security domains.

**Acknowledgements:** Work is supported in part by: DOE grant DE-SC0017810, NSF grant NSF-PHY-1914059, the University of Notre Dame – through the Resilience and Recovery Grant Program, and QuarkNet for HS Teacher and Student support.**References:**
[1] "Basic Research Needs for High Energy Physics Detector Research & Development," https://science.osti.gov/-/media/hep/pdf/Reports/2020/DOE_Basic_Research_Needs_Study_on_High_Energy_Physics.pdf

[2] R. Ruchti, et al, "Advanced Optical Instrumentation for Ultra-compact, Radiation Hard EM Calorimetry", LOI submitted to the Snowmass 2021 Instrumentation and Energy Frontiers, https://www.snowmass21.org/docs/files/summaries/IF/SNOWMASS21-IF6_IF4-EF1_EF4-102.pdf

[3] M. Aleksa, et al, "Calorimeters for the FCC-hh", CERN-FCC-PHYS-2019-0003.

[4] A. Ledovskoy, "Shashlik Timing", Report to the RADiCAL Group, January 17, 2021, https://notredame.box.com/s/vjpbvpxn2sff72y6y4wvue3nnhcucgtx

[5] A. Ledovskoy, "RADiCAL Studies", Report to the RADiCAL Group, September 12, 2021, https://notredame.box.com/s/p9afzt9hxp2yg22vmoycfir8fnnt7tzk

[6] "The Shashlik Calorimeter, a LYSO/W Plate Calorimeter for Precision EM Calorimetry in the High Luminosity LHC Environment", B. Cox, 38th International Conference on High Energy Physics Chicago (2016) http://indico.cern.ch/event/432527/contributions/1071731/

[7] https://cms.cern/detector/measuring-energy/crystal-calorimeter

[8] "Shashlik Performance", A. Ledovskoy, Presentation to the CMS Endcap Calorimetry Review, 10 December 2014, https://www3.nd.edu/~rruchti/Shashlik_Performance.pdf

[9] "Event Display for VBF Jets in Shashlik", A. Ledovskoy, Presented in the Shashlik + Enhanced HE Meeting (March 11, 2015) https://www3.nd.edu/~rruchti/Event_Display_For_Jets.pdf

[10] "LYSO based precision timing calorimeters", A. Bornheim, S. Xie, J. Duarte, M. Spiropulu, J. Trevor, D. Anderson, A. Ronzhin, A. Apresyan, C. Pena, and M. Hassanshahi, Submitted to the Proceedings of CALOR 2016, Daegu, Republic of Korea (15-20 May 2016) https://indico.cern.ch/event/472938/sessions/102269/#20160517

[11] "Advanced Optical Instrumentation for Ultra-compact, Radiation Hard, Fast-timing Calorimetry", R. Ruchti et al, Presentation at CPAD2021, Stony Brook University, March 18, 2021, https://indico.fnal.gov/event/46746/contributions/210058/

[12] "RADiCAL – "Radiation-hard Innovative EM Calorimetry", R. Ruchti et al, Presentation at DPF2021, Florida State University, July 12, 2021. https://indico.cern.ch/event/1034469/contributions/4431681/20